\documentclass{article}

\usepackage{arxiv}
\usepackage{amsmath}
\usepackage[utf8]{inputenc} 
\usepackage[T1]{fontenc}    
\usepackage{cite} 
\usepackage[hyperfootnotes=false,colorlinks=true,linkcolor=black,anchorcolor=black,citecolor=black,filecolor=black,menucolor=black,runcolor=black,urlcolor=black]{hyperref}       
\usepackage{url}            
\usepackage{booktabs}       
\usepackage{amsfonts}       
\usepackage{nicefrac}       
\usepackage{microtype}      
\usepackage{amssymb}
\usepackage{float}
\usepackage{listings}
\newsavebox{\LstBox}
\usepackage{graphicx}
\usepackage{gensymb} 
\usepackage{caption}
\usepackage[round]{natbib}
\usepackage{doi}
\usepackage{nopageno}
\usepackage{enumitem}
\usepackage[OT2,OT1]{fontenc}
\newcommand\cyr
{
\renewcommand\rmdefault{wncyr}
\renewcommand\sfdefault{wncyss}
\renewcommand\encodingdefault{OT2}
\normalfont
\selectfont
}
\DeclareTextFontCommand{\textcyr}{\cyr}

\setlist[itemize]{leftmargin=*}

\defcitealias{inebriati}{Mitchell, Webb et al, 2010}
\title{The Ballmer Peak: An Empirical Search}
\date{}

\author{ \href{https://orcid.org/0009-0003-3567-2181}{\includegraphics[scale=0.06]{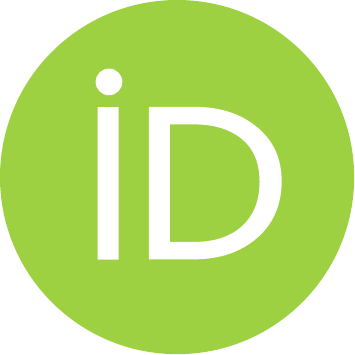}\hspace{1mm}Twm~Stone}\\
	A Real Job\\
	The Private Sector\\
	Cambridge, UK \\
	\texttt{twm.stone@cantab.net} \\
	\And
	\href{https://orcid.org/0000-0002-1834-8705}{\includegraphics[scale=0.06]{orcid.pdf}\hspace{1mm}Jaz~Stoddart} \\
	A Real Scientist\\
	Royal Botanic Gardens, Kew\\
	London, UK\\
	\texttt{js2231@cantab.ac.uk} \\
}

\renewcommand{\headeright}{Case Study}
\renewcommand{\undertitle}{Clinical Study}
\hypersetup{
pdftitle={The Ballmer Peak: An Empirical Search},
pdfsubject={cs.SE},
pdfauthor={Twm~Stone, Jaz~Stoddart},
}

\begin{document}
\begin{lrbox}{\LstBox}
\begin{lstlisting}[language=Python, basicstyle=\footnotesize]
def mostFrequentEven(self, nums: List[int]):
  evens = sorted([elem for elem in nums
     if elem % 2 == 0])
  counts = {thing: evens.count(thing) for
     thing in evens}
  return -1 if not counts else sorted([elem
     for elem in counts if counts[elem] ==
     max([counts[thing] for thing in
     counts])])[0]

\end{lstlisting}
\end{lrbox}

\twocolumn[ 
  \begin{@twocolumnfalse} 
  
\maketitle
\vspace{-20pt}
\begin{abstract}
The concept of a `Ballmer Peak' was first proposed in 2007, postulating that there exists a very specific blood alcohol content which confers superhuman programming ability. More generally, there is a commonly held belief among software engineers that coding is easier and more productive after a few drinks. Using the industry standard for assessment of coding ability, we conducted a search for such a peak and more generally investigated the effect of different amounts of alcohol on performance. We conclusively  refute the existence of a specific peak with large magnitude, but with $p<0.001$ find that there was a significant positive effect to a low amount of alcohol---slightly less than two drinks---on programming ability.
\end{abstract}

\vspace{0.45cm}

  \end{@twocolumnfalse} 
] 

\textbf{ACM Reference format}:\newline
Twm Stone and Jaz Stoddart. 2024. The Ballmer Peak: An Empirical Search. In \textit{Proceedings of SIGBOVIK, \mbox{Pittsburgh}, PA
USA, April 2024 (SIGBOVIK’24)}, 7 pages.

\section{Introduction}
There has long been a belief among programmers that their coding ability is significantly improved after a couple of drinks. Although there has been significant previous work in this area---notably showing a minor beneficial impact of alcohol on creative problem solving \citep{creativity}, and showing a detrimental impact of a particular (high) level of inebriation on novice coders \citep{jss}---there has not been any direct scientific investigation of the effect of differing levels of alcohol intoxication on coding ability.

It was posited in 2007 by renowned popular science author Randall Munroe, in the adjacent comic \citep{xkcd}, that there existed a peak of width approximately 0.01\%, centred on $0.1337\%$ blood alcohol content, which confers superhuman programming ability. More recently the intrepid comedic minds of Mitchell, Webb et al. explored the huge potential benefits to all human activities of having a very precise level of inebriation (slightly less than two drinks \citepalias{inebriati}) and an insightful Danish-language documentary by Vinterberg demonstrated the wide-ranging positive effects of maintaining a blood alcohol level of above 0.050\% \citep{druk} .
\begin{figure}[h]
\centering
\includegraphics[width=0.5\textwidth]{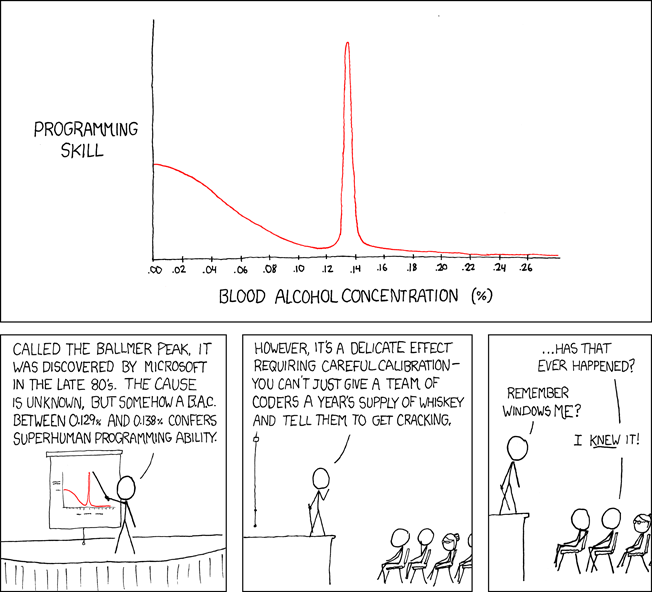}
\caption{\textit{Apple uses automated schnapps IVs.}}
\label{XKCD}
\end{figure}
\captionsetup[figure]{justification=centering}

\vspace{-20pt}
In this study, we attempt to establish the existence of the Ballmer Peak and more generally the existence, magnitude and location of any benefit to programming conferred by alcohol intoxication. We additionally aim to gain a qualitative understanding of the different ways in which alcohol consumption impacts the different facets of programming, and to guide future research in this area.

\section{Methodology}
\subsection{Test subject}
The test subject was chosen so that they had substantial previous experience in both competitive programming and using the chosen programming language professionally, but had never used the LeetCode platform before. This minimized the impact of the practice effect on problem completion rate whilst also preserving the novelty of the problems.

The selected candidate was a male, 27 year old Caucasian of Northern European descent, currently employed as a Software Engineer. Weighing around 65 kilograms and being approximately 170cm tall, he is generally fit and well with no formally diagnosed cognitive or physical conditions which might impact problem solving ability or result in abnormal metabolism of alcohol.\footnote{Due to requirements of the Ethics Board, the test subject was the first author.}

\subsection{Means of intoxication}
To achieve a given level of percentage blood alcohol content (\%BAC) we needed to administer potentially large amounts of alcohol to the test subject but in a predictable manner. Mitchell, Webb et al. suggested several options for administration---Cointreau enemas, intravenous claret, enteral Special Brew---which were deemed infeasible by the Ethics Board due to lack of available medical expertise \citepalias{inebriati}.

Frequent ingestion of liqueur chocolates was considered as the method of intoxication but the required quantities and rate of consumption was deemed financially impractical and liable to induce emesis. It was thus decided that the most cost-effective means of administering ethanol was to imbibe it. This was done with a combination of beer, cider, premixed cocktails, and something which could generously be called `homemade punch', the proportions of which depended on the target \%BAC. These dilute sources of ethanol were chosen as pure ethanol is quite unpalatable, far more expensive to purchase and infeasible for us to produce ourselves, owing to the difficulty associated with separation from the azeotrope it forms with water.

\subsection{Assessment of coding ability}
To assess coding ability in this investigation, we opted to use the industry standard---the speed of solving LeetCode problems.\footnote{\texttt{\href{https://leetcode.com/}{https://leetcode.com/}}} The chosen language for solving the LeetCode problems was a relatively new one---akin to pseudo-pseudocode with duck typing---introduced by Van Rossum and Drake, since it was the language with which the test subject had the greatest familiarity \citep{python}. The LeetCode problems were filtered using inbuilt functionality to select at random exclusively from the subset of problems ranked as "Easy"; problems requiring a subscription or the use of other languages were also discarded, as were problems previously completed.\footnote{Also any problem involving binary trees on account of the test subject not being bothered to learn how they work.} The problems were solved by typing directly into the provided in-browser IDE; no code completion, debugging tools or AI assistance were used, although it did have syntax highlighting.\footnote{This was mildly useful on a number of occasions, although it did not prevent the test subject spending 10+ minutes failing to realise he had misspelled \texttt{lambda} as \texttt{lamdba}.}

Since our budget did not extend to a premium LeetCode subscription, we did not use the associated debugger but instead simulated the problem-solving workflow of a professional software engineer by
\begin{itemize}
\item Adding lots of \texttt{print()} statements, running all of the provided test cases, and looking at \texttt{stdout}.
\item Getting annoyed and looking up how various language features actually work on Stack Exchange.
\end{itemize}

The time to solve was taken from the point the page loaded to the point the submission stats loaded for a correct solution, including all reading, coding, debugging, and run time. No attempt to measure the asymptotic behaviour of the solution---in particular, as long as the platform accepted a solution and did not return a Time Limit Exceeded error, it was considered solved even if the algorithmic complexity was not as good as required in the problem description---and no other measure of code `quality' was quantified, although observations were made by the author contemporaneously.\footnote{We originally used Microsoft Word for this but it kept putting the entire document into 8pt Times New Roman whether we wanted it to or not.}

\subsection{Recording of blood alcohol content}
Blood alcohol content (\%BAC) was recorded using a BACtrack C6 electronic breathalyser,\footnote{BACtrack Breathalyzers / KHN Solutions Inc., San Francisco, USA} which was recently calibrated and claims an accuracy of $\pm0.001\%$BAC. The accuracy of the breathalyzer was verified post-calibration through multiple means.

First, the zero point reading was verified through testing on sober volunteers.\footnote{Here sober is defined as no alcohol having been consumed within the preceding 24 hours.} Repeated readings taken on them consistently returned values of 0.000\%BAC.\footnote{It was observed during later testing that it always rounded down readings below 0.007\% which contradicts the manufacturer claimed precision.}
Four sober volunteers were then each given several pints of approximately equal strength beer or cider (Fuller's London Pride, 4.7\% or Aspall Cyder, 4.5\%)\footnote{Due to availability at The Old Wheatsheaf, Enfield.} and periodically breathalysed until a peak value was reached.\footnote{They were all former university students and thus experienced drinkers, mitigating any risk factors that may have been grounds for ethical concerns.} The expected peak \%BAC for each volunteer was calculated from known volume of consumed alcohol and expected blood volume using Equation 1, which roughly matched the observations taken using the breathalyser.

\[\%BAC = 100 \times \frac{V_{alcohol}}{V_{blood}} = 100 \times \frac{V_{alcohol}}{M_{patient} * B} \tag{1}\]

where $V_{alcohol}$ is the volume of alcohol consumed in ml, $V_{blood}$ is the volume of blood in ml, $M_{patient}$ is the mass of the patient in kg, and B is the gender and age appropriate value for the average blood volume in $mL.kg^{-1}$ \citep{blood}.\footnote{B for an adult male = 75$mL.kg^{-1}$ and for an adult female = 65$mL.kg^{-1}$}

Repeated readings taken from the same subject had a spread of $\pm0.002\%$, in line with the advertised accuracy of the breathalyzer.

\subsection{Data collection}
Data collection took place approximately every second night for forty days and forty nights. The test subject was inebriated to a given level and then, after a pause to minimise residual alcohol on the breath,\footnote{Including efforts to wash the mouth out with water.} began solving problems. The \%BAC of the test subject was monitored through measurements taken immediately before and after each problem was attempted. After the subject became bored with coding,\footnote{Or was observed buying crisps, or trying to remember the Oscar-winning films of the 70s.} he resumed drinking and the process was repeated up to several times.

\section{Results}
A total of 100 problems were attempted\footnote{Of the 100, 99 problems were completed and 1, attempted at $\sim$0.2\%BAC, was not owing to the the test subject succumbing to unconsciousness after half an hour. The recorded completion time of 90 minutes was projected based on a sober assessment of the problem and the progress made.} over the course of forty days and forty nights preceding the SIGBOVIK '24 conference, on 15 separate occasions. The results were distributed as follows. Note that there is a bias towards data collection in the lower ranges of \%BAC; owing to the average time per problem being greatly reduced more were completed before more drinks were consumed.
\begin{table}[H]
	\caption{Sample distribution}
	\centering
	\begin{tabular}{crr}
		\toprule
		Range     & Sample size     & Mean time /s \\
		\midrule
		$x =$ 0.000\% & 22  & 394    \\
		0.000\% $< x \leqslant$  0.025\%     & 16 & 402      \\
		0.025\% $< x \leqslant$  0.050\%     & 20 & 249      \\
		0.050\% $< x \leqslant$  0.075\%     & 12 & 310      \\
		0.075\% $< x \leqslant$  0.100\%     & 11 & 610      \\
		0.100\% $< x \leqslant$  0.125\%     & 6 & 614      \\
		0.125\% $< x \leqslant$  0.150\%     & 8 & 1204      \\
		0.150\% $< x \leqslant$  0.175\%     & 3 & 2550      \\
		0.175\% $< x \leqslant$  0.200\%     & 2 & 3281      \\
		\bottomrule
	\end{tabular}
	\label{tab:table}
\end{table}

Our first figure, below, is a scatter plot of the raw data---showing completion time against \%BAC.

\begin{figure}[H]
\centering
\includegraphics[width=0.5\textwidth]{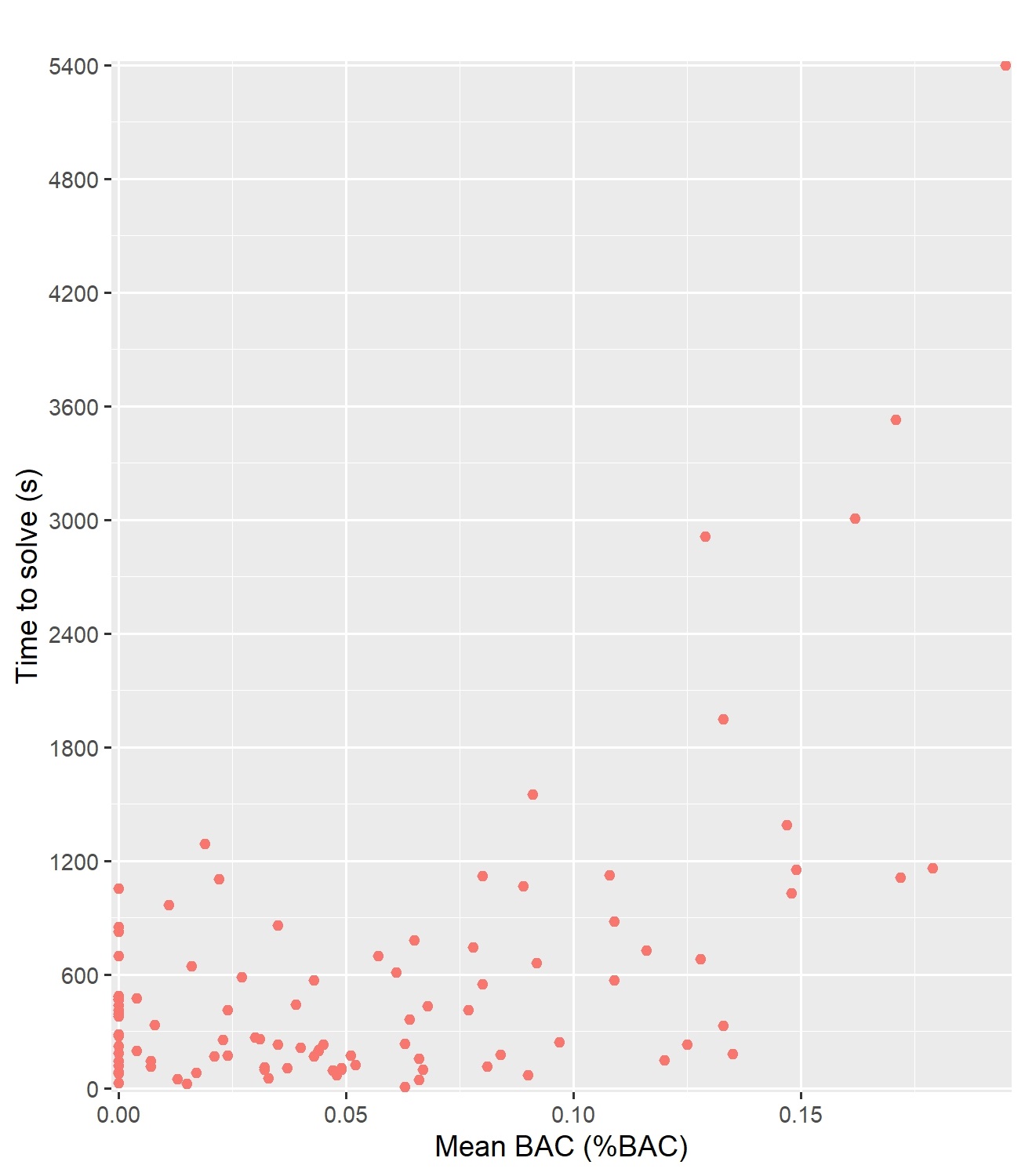}
\caption{Time to solve in seconds against mean \%BAC.}
\label{raw}
\end{figure}
\captionsetup[figure]{justification=centering}

\begin{figure}[H]
\centering
\includegraphics[width=0.5\textwidth]{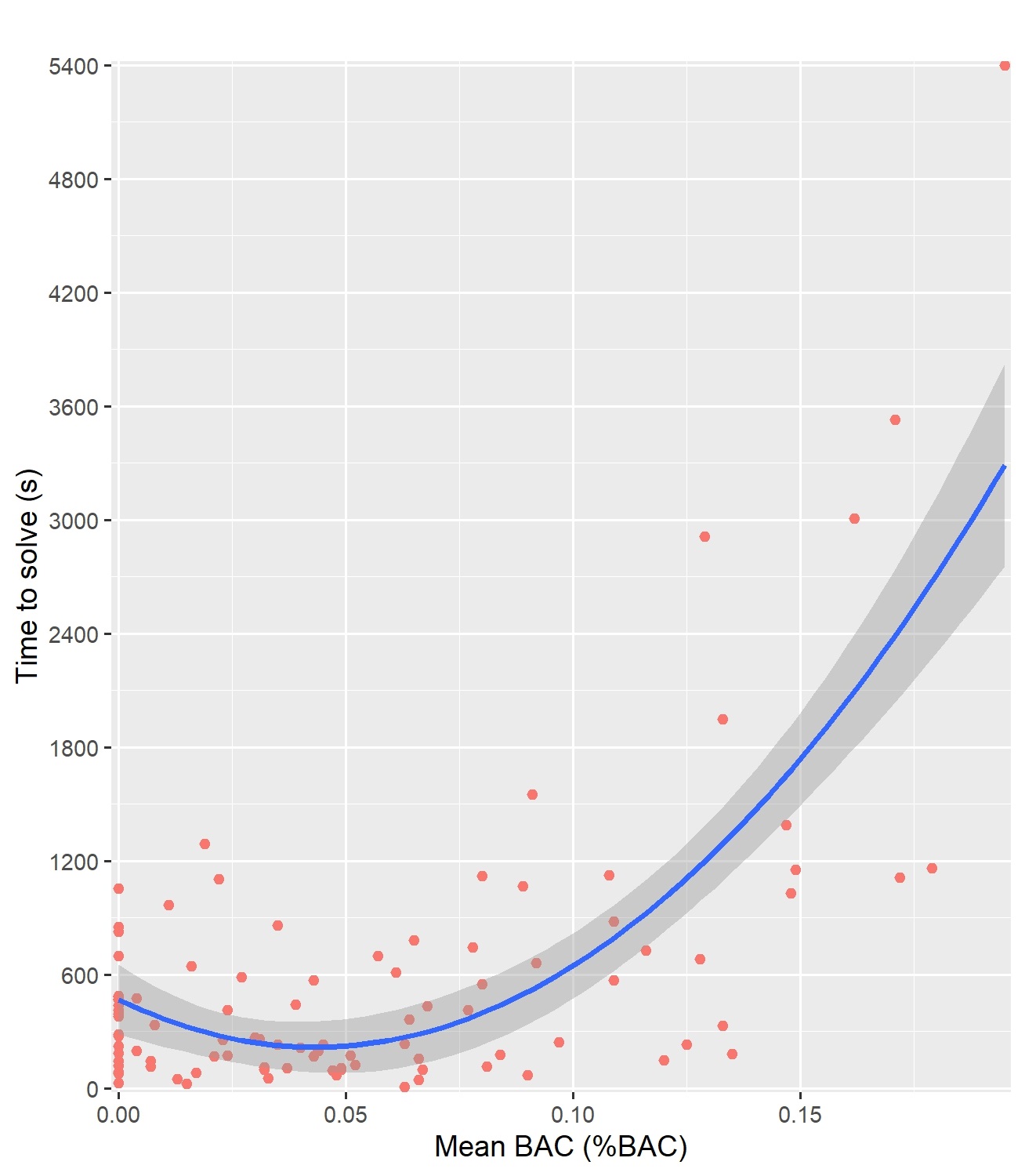}
\caption{Time to solve against mean \%BAC.}
\label{rawLOBF}
\end{figure}

\begin{figure}[H]
\centering
\includegraphics[width=0.5\textwidth]{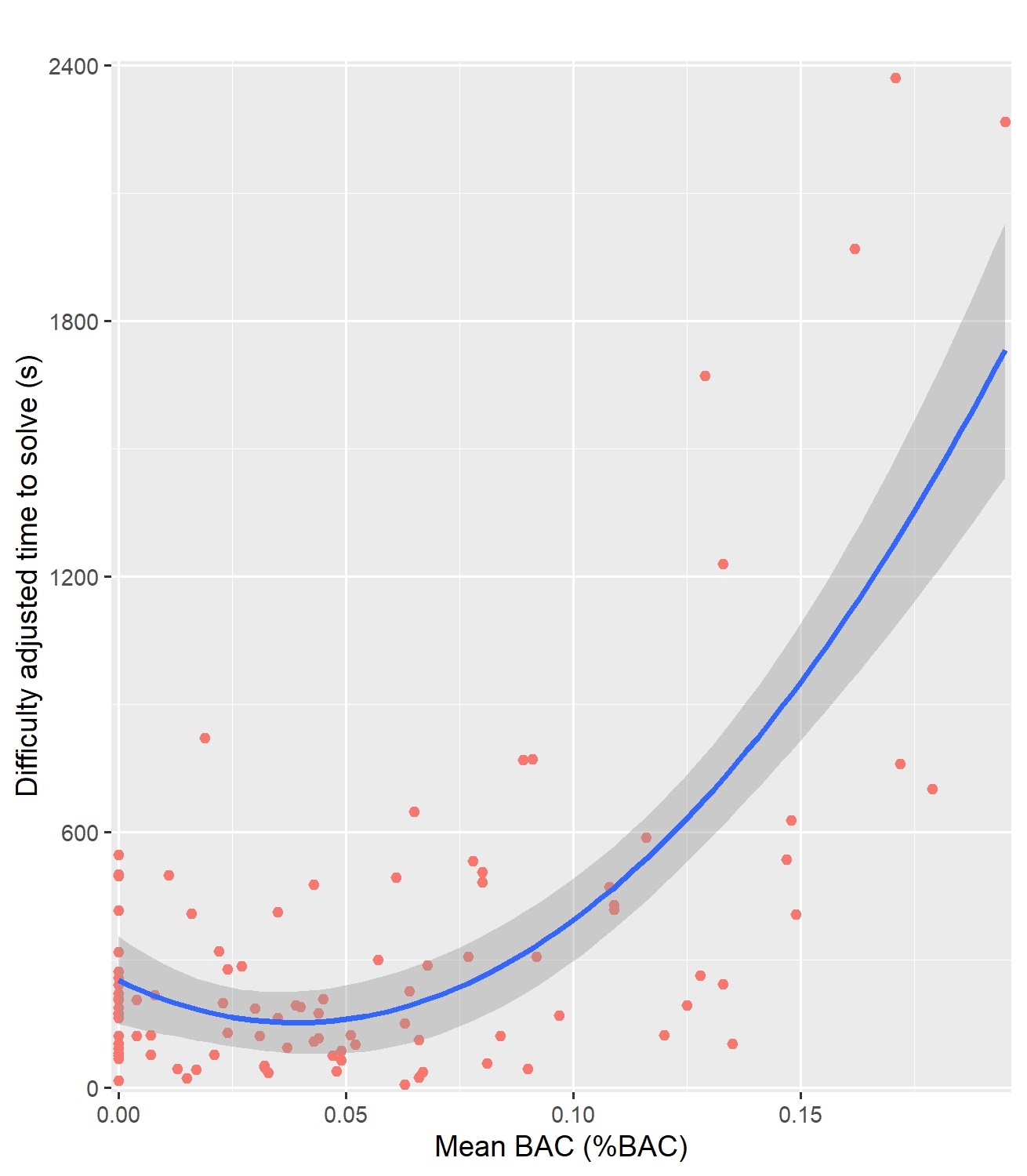}
\caption{Difficulty-adjusted time against mean \%BAC.}
\label{adjustedLOBF}
\end{figure}

The next two figures presents the same data, but with a quadratic line of best fit and a 95\% confidence ribbon. This model was chosen from many potential ones (linear, quadratic, exponential, cubic, power, etc.) since it matched most closely; it is fit with $p < 0.001$. Figure~\ref{rawLOBF} presents the raw data and has a minimum at (0.043\%, 222s) and y-intercept (0.000\%, 466s), whereas Figure~\ref{adjustedLOBF} presents "adjusted time" (see \nameref{adjust}) against \%BAC in the same way, with a minimum at (0.047\%, 64.5s) and y-intercept (0.000\%, 214s).

\captionsetup[figure]{justification=centering}

\captionsetup[figure]{justification=centering}

\begin{figure}[H]
\centering
\includegraphics[width=0.5\textwidth]{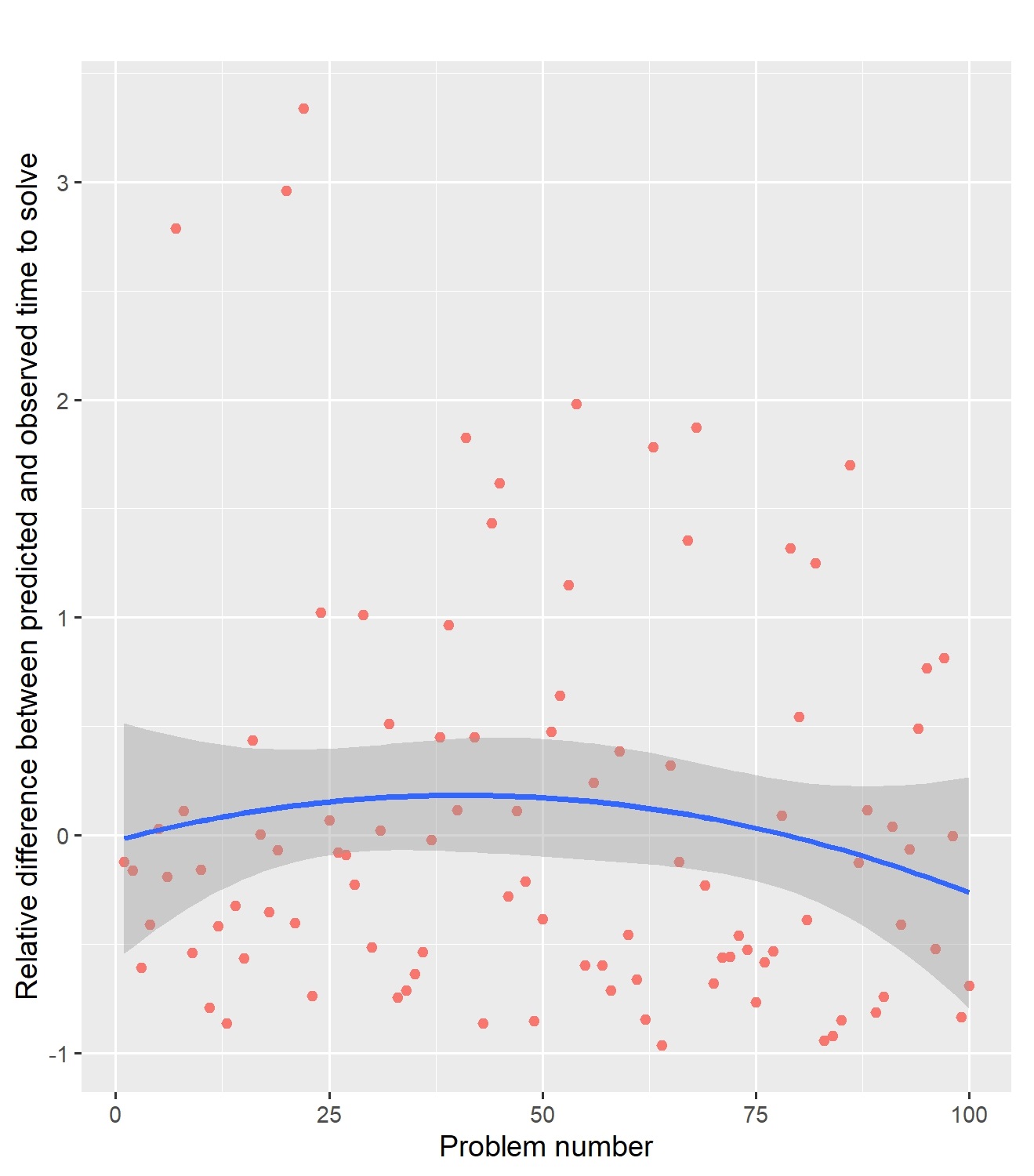}
\caption{Prediction error throughout the study.}
\label{practice}
\end{figure}

In order to visualise whether there was a "practice effect"---where performance improved over time across all problems irrespective of intoxication---Figure~\ref{practice} was produced. The relative difference between observed and predicted time to solve was a metric calculated to attempt to account for \%BAC and thus allow for trends in performance over the problem set to be seen.
The relative difference between observed and predicted time was calculated as:

\[\textit{Difference}_{relative} = \frac{T_{observed} - T_{predicted}}{T_{predicted}} \tag{2}\]

\captionsetup[figure]{justification=centering}

\section{Discussion}
\subsection{Results and Implications}
Over the course of this study the test subject descended through, at least once, every possible \%BAC between 0.000\% and 0.200\%. Despite this, no spike in perceptual or actual competence was observed from problems suddenly becoming easier at any given level of inebriation. Thus, we believe we can conclusively rule out the existence of the Ballmer Peak as originally formulated.

However, looking at Figure~\ref{rawLOBF} it is clear that a moderate amount of alcohol has positive effects on problem-solving speed. The peak of this effect is at about 0.043\%BAC, where the test subject was able to solve problems around 45\% faster than while sober. To get to this level requires slightly less than two standard drinks---that is, around a pint and a half of 5\% beer, a glass and a half of rosé, or three shots. Additional alcohol intake beyond that consistently worsened performance, with only slightly higher \%BAC having parity with sobriety and then higher levels having an increasingly negative effect. After the initial peak there was no positive impact at any point for a marginal drink.

A further observation is that alcohol intake increases spread but does not necessarily lower the `best case' solving speed. One of the fastest problems solved during this study was at 0.090\%BAC, the equivalent of having 6 straight shots of vodka.\footnote{This clearly had nothing to do with the fact that the problem in question---return true if there are three consecutive odd numbers in this integer array---was orders of magnitude simpler than some of the other problems in the so-called `Easy' category.} This is in accordance with the notes taken contemporaneously; there was a perception that sometimes one can just "see it" and get the solution straight away; this ability did not actually seem to be affected by alcohol intake less than a very high level. However, as soon as the problem required debugging or trying a different approach the ability of the test subject regressed significantly. This was due to a combination of continual typographic and logical errors---each requiring some amount of effort to fix, the fix itself potentially adding more---but also the test subject remained convinced for longer that `this approach is basically perfect I just need to fiddle with the algorithm a bit' rather than trying a different approach. 

For the sober problems the test subject made more of an attempt to use so-called `idiomatic Python' in the hope of getting a pithy one-line solution as fast as possible. This was sometimes successful but often wasted a lot of time fiddling with list comprehensions\footnote{For example, the solution for \href{https://leetcode.com/problems/most-frequent-even-element/}{2404: Most Frequent Even Element} was the following concise yet readable code: \par\usebox{\LstBox}} although it did reliably place him in the top 95\% of submissions (see below).

\begin{figure}[h]
\centering
\includegraphics[width=0.5\textwidth]{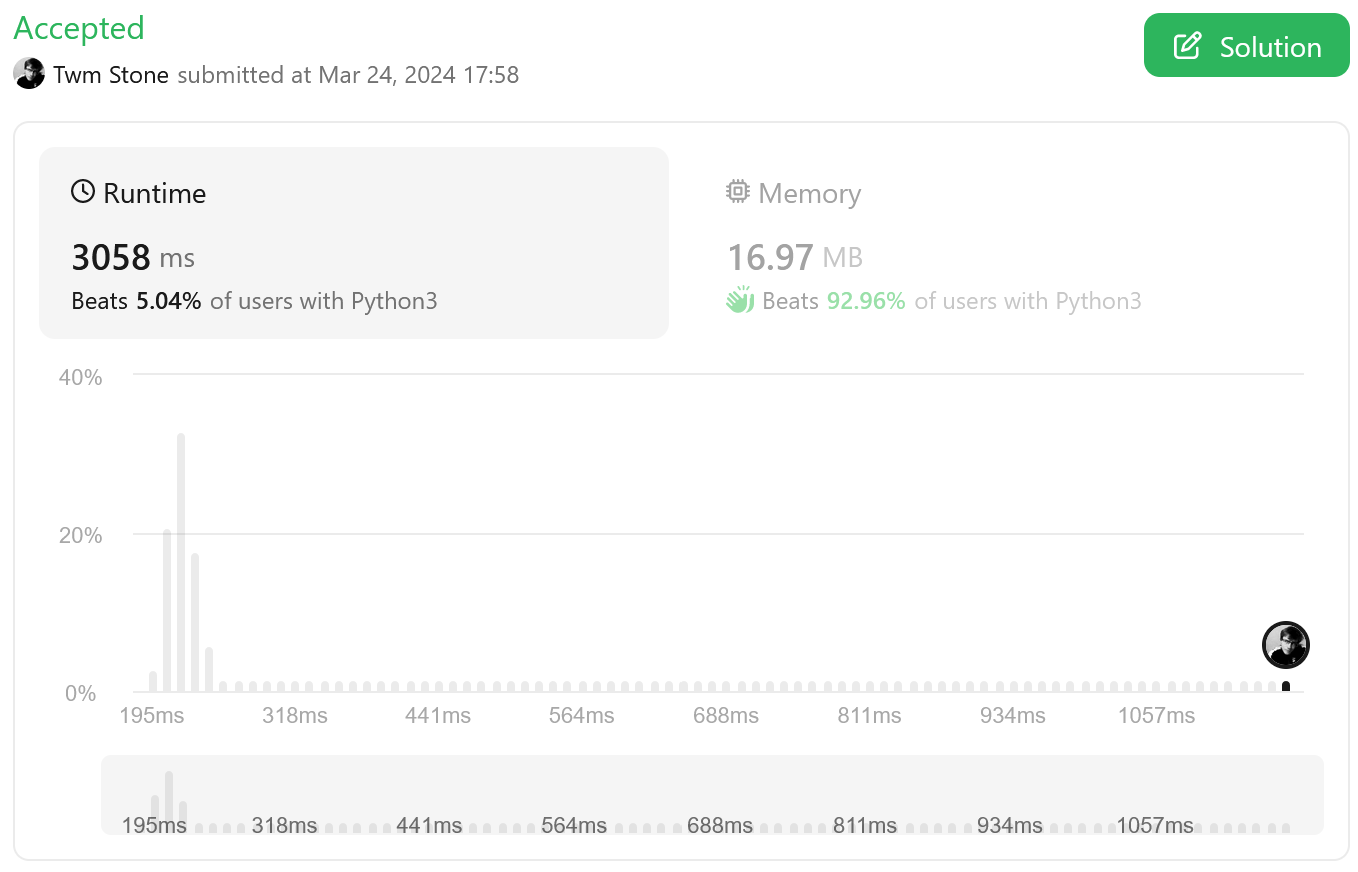}
\caption{Somehow, 5\% of users did worse than this...\protect\footnotemark}
\end{figure}
\captionsetup[figure]{}

The possibility of a practice effect was considered. Once a line of best fit had been determined for the full data set, we measured the deviation from this `predicted time` for each of the problems solved. The results, Figure~\ref{practice}, of this analysis suggest there is no significant practice effect in our dataset. This conclusion is reached given the best fit for this data was a quadratic curve rather than a line with negative gradient. Additionally, the confidence ribbon entirely overlaps the line $y=0$; indicating the observed values never significantly deviate from predicted values made with a model that assumes no practice effect.

\vspace{-1.21pt}
\subsection{Limitations}
\footnotetext{Astute readers will have noticed that the x-axis labels are not consistent with the achieved runtime---in other words, this result was off-the-charts bad.}\label{adjust}Assessing how long a problem `should' have taken was difficult; LeetCode does not provide average solution time for its problems, only `acceptance rate', i.e. what proportion of submissions for the problem were correct. This was used as a proxy for difficulty in our `adjusted time' but might reasonably also be considered to reflect how many traps or tricks there were in the question. We directly divided the real completion time by the acceptance\% to adjust but one might be sceptical that is realistic---it is improbable that "add two numbers together" (\#2235, 87.6\%) is really half the difficulty of "find a substring length $m$ which repeats $k$ times" (\#1566, 43.0\%).\footnote{In particular, this second problem requires not only adding but also counting, and we know from \citep{difficult} that, for the superstitious, the combination of the two requires dealing with both cardinals and ordinals.} Figure~\ref{adjustedLOBF} shows the impact of adjusting for difficulty in this manner. The location of the peak is almost unchanged and the overall data distribution is similar. Alternatives were considered for measuring difficulty but getting sober volunteers to solve enough of the problem set to have meaningful results on this was deemed infeasible within the time constraints of the study.
\vspace{-1pt}

The approach to measuring \%BAC used during this study may have impacted the accuracy of results. It should be noted that whilst it might have been possible to complete this work without the assistance of a breathalyser, using estimates based upon Equation 1, the use of one is likely to have increased accuracy and ease with which \%BAC  was measured. Estimates based solely upon blood volume are subject to introduced errors from assumptions made regarding rates of metabolism, efficacy of intake alcohol into blood, body fat distribution and proportion, and, of course, blood volume \citep{liver} \citep{blood}. However, breathalyzers \textit{are} known to over estimate \%BAC immediately following consumption and so this may present a further source of inaccuracy---though measures were taken to mitigate this, such as waiting between initial consumption and starting the test and washing the mouth out with water.

Alternative more accurate approaches to \%BAC measurement were not considered due to the associated prohibitive costs, impracticality, and ethical issues presented by repeated frequent sampling of blood and urine.

\subsection{Further confounders}
This study was not double-blinded or indeed blinded at all. The test subject knew\footnote{At least, if they were paying attention.} how much alcohol they had consumed. Even worse, given that they had taken the measurement of \%BAC themselves and were aware of previous results, there may have been a placebo effect around how hard the test subject expected the problem to be, which could have informed their approach.

Increased tolerance of the subject to alcohol over the course of the study was not accounted for. Whilst initial data gathering was performed after pre-existing social events / pub trips, it was quite quickly established that this level of drinking was much too slow to realistically reach above $\sim$0.08\%BAC. In order to get data for higher ranges (including, crucially, the originally postulated 0.1337\% peak) the test subject had to intentionally drink large quantities rapidly on multiple occasions. This was observed to have a significant positive effect\footnote{From a certain point of view...} on his ability to withstand high levels of inebriation during subsequent research.\footnote{At least perceptually, their ability to focus without feeling ill was improved substantially. The fact that that the test subject ran a marathon mid-way through the study and then ceased training may have also impacted this.}

Tiredness and caffeine intake may have influenced the ability of the test subject to solve the problems. Whilst the test subject was never knowingly under-caffeinated some of the problems were solved quite late at night. The test subject's sleep cycle \textit{did} suffer adverse consequences from this research---hence much of the data was gathered on Friday or Saturday nights to minimize the impact on their day job---and it is possible data gathered towards the end of the study was impacted negatively by this.

Prolonged alcohol consumption followed by periods of unconsciousness in which one does not drink water causes changes in composition of the blood. Of these changes, it is the reduced volume from dehydration and the presence of alcohol metabolites---primarily  acetaldehyde, as well as additional congeners such as tannins and other phenolic compounds---which are believed to be the principal causes of a hangover \citep{hangover}. Given the nature of the study and the concentrated periods of data collection throughout the forty days and forty nights of study, it is possible that the test subject achieved a state of both being drunk and hungover at the same time. Since this was not a study aimed at examining the effects of programming while hungover, we chose to ignore any impacts this might have had.

\section{Conclusions}
From this brief case study, we believe there is sufficient evidence to conclude that the Ballmer peak, sensu Munroe (2007), does not exist. The absence of a specific narrow peak for improved programming ability does not however discount the more widely held beliefs of general improvements in performance after very nearly two drinks.

To that end, our work supports the hypotheses of Mitchell, Webb et al. and we believe that the high importance of the subject matter means that further work to replicate this study on a larger scale is thus needed.\footnote{Additionally, we currently lack understanding as to the mechanistic, psychological and metabolic processes which have led to this result. There are large potential benefits to the field of software engineering from improving our knowledge of how to exploit this effect.} 

Cheers.

\subsection{Additional conclusions}
In addition to our primary areas of research, we discovered various other items of scientific value:
\begin{itemize}
\item Recording data for research while becoming progressively more inebriated is challenging.\footnote{See \href{http://bitly.com/98K8eH}{bit.ly/98K8eH} for more evidence on this.}
\item The UK drink-driving limit is dangerously high. The test subject could literally\footnote{Although did not!} five-and-drive---drink 5 beers over the course of an evening and reasonably expect to be under the legal limit while leaving the pub. This is not to say that the test subject could reasonably expect to be safe to drive in this condition, just that it would be legal.
\item It was learned that for the dial in a fridge "2" means "2 power" and not "2 \degree C" and the fridge will happily chug along at 6+ \degree C. Furthermore milk stored at 6+ \degree C will spoil and spoiled milk actually isn't okay to drink. Unintentionally drinking several glasses of slightly off milk will make you rather ill.\footnote{Spoiled milk typically contains some small amount of alcohol and thus there may be grounds to return to the study of spoiled milk as a low concentration source of alcohol for maintenance of \%BAC at the optimal level. Products such as Kumis contain mild levels of alcohol and there are vodkas now distilled purely from milk based alcohol.}
\end{itemize}

\subsection{Suggestions for further research}
We have identified several promising avenues for continuing research in this area. It seems likely that the optimal level of alcohol consumption is different for design, coding, testing, debugging, documentation, etc. All of these could be interesting areas to explore.

The immediate applicability of this research to software engineering is also potentially limited, since professional coding work is likely to require working with a larger codebase, not having the entire context for the problem in your head, and negotiating with stakeholders. Each of these could favourably or adversely be impacted by alcohol intake.

Auto-brewery syndrome is a rare condition in which stomach flora can produce a constant supply of alcohol keeping the afflicted individual in a permanent state of intoxication \citep{abs}. Further work may be required to induce auto-brewery syndrome to see if one can achieve a constant level of inebriation sufficient to exploit the findings of this work and improve programming performance.

Further research will commence as soon as we secure the funding to replenish Twm's drinks cabinet.

\section{Acknowledgements}
We would like to thank the following people, without whom this paper would have been slightly worse:

\begin{itemize}

\item Amanda Chua and Kevin Lim for procuring supplies, for helping with data collection and for advice on our experimental design.

\item Our Ethics Board, Tom Flynn and Rachel Newhouse, for rubber-stamping all of our dubious decisions.

\item Joonatan Laulainen and Dan Rob\hspace{-0.34cm}{\cyr e}rtson, for steering us clear from even worse ideas for a study, for convivial ideation and for moral support.

\item N\hspace{-0.49cm}{\cyr e}phele P\hspace{-0.49cm}{\cyr e}nny, Jack Rickard and Oliver Shenton for assisting in the measurements of calibration accuracy of our equipment.

\item Clara Ding, Tim Coulter, Lily Mills, and Dmitri \mbox{Whitmore}, for putting up with Twm drinking for this study while taking part in a video game tournament.

\item Betty la Chatte for guidance on creative typography.

\item Herbie Bradley, Patrick Kennedy-Hunt and Alexandra Souly for proof-reading and advisement.
\end{itemize}

\bibliographystyle{unsrtnat}
\bibliography{references}
\end{document}